# Reconstructing solar wind inhomogeneous structures from stereoscopic observations in white-light: Small transients along the Sun-Earth line

Xiaolei Li[1,2], Yuming Wang[1,3,4,*], Rui Liu[1,2,4], Chenglong Shen[1,2,4], Quanhao Zhang[1,2], Bin Zhuang[1,2], Jiajia Liu[5], and Yutian Chi[1,4]

[1]CAS Key Laboratory of Geospace Environment, Department of Geophysics and Planetary Sciences, University of Science and Technology of China, Hefei, China
[2]Collaborative Innovation Center of Astronautical Science and Technology, Hefei 230026, China
[3]Synergetic Innovation Center of Quantum Information & Quantum Physics, University of Science and Technology of China, Hefei 230026, China
[4]Mengcheng National Geophysical Observatory, University of Science and Technology of China
[5]Solar Physics and Space Plasma Research Center, School of Mathematics and Statistics, University of Sheffield, Sheffield S37RH, UK
*Corresponding author, ymwang@ustc.edu.cn

**Abstract** The Heliospheric Imagers (HI) on board the two spacecraft of the Solar Terrestrial Relations Observatory (STEREO) provided white-light images of transients in the solar wind from dual perspectives from 2007 to 2014. In this paper, we develop a new method to identify and locate the transients automatically from simultaneous images from the two inner telescopes, known as HI-1, based on a correlation analysis. Correlation coefficient (cc) maps along the Sun-Earth line are constructed for the period from 1 Jan 2010 to 28 Feb 2011. From the maps, transients propagating along the Sun-Earth line are identified, and a 27-day periodic pattern is revealed, especially for small-scale transients. Such a periodicity in the transient pattern is consistent with the rotation of the Sun's global magnetic structure and the periodic crossing of the streamer structures and slow solar wind across the Sun-Earth line, and this substantiates the reliability of our method and the high degree of association between the small-scale transients of the slow solar wind and the coronal streamers. Besides, it is suggested by the cc map that small-scale transients along the Sun-Earth line are more frequent than large-scale transients by a factor of at least 2, and that they quickly diffused into background solar wind within about 40 Rs in terms of the signal-to-noise ratio of white-light emissions. The method provides a new tool to reconstruct inhomogeneous structures in the heliosphere from multiple perspectives.

1. Introduction

It is widely accepted that fast solar wind (typically > 500 km s$^{-1}$) originates from large coronal holes (e.g., Hollweg and Isenberg 2002), while slow solar wind (< 500 km s$^{-1}$) from boundaries of corona holes or from closed coronal loops surrounding the heliospheric current sheets (HCS) and streamer belts (e.g., Feldman & Widing 2003; Kasper et al., 2007, 2012; Stakhiv et al., 2015, 2016).

Inhomogeneous structures, contrasting with continuum, are important components of the solar wind, including transients, waves and turbulence. Compared with the large-scale transients in the solar wind, such as collisionless interplanetary shock waves and coronal mass ejections (CMEs), small-scale solar wind transients looking like 'blobs' or 'ripples' are much more common. They mostly appear to emerge from coronal streamers as clear density enhancements but eventually diffuse to become part of the slow solar wind (e.g., Wang et al., 1998; Rouillard et al., 2010; Viall et al., 2010; Viall & Vourlidas 2015; Plotnikov et al., 2016; Sanchez-Diaz et al., 2017).

The solar wind can be detected locally by in-situ instruments, but such instruments cannot give a global picture of the solar wind. Owing to the density fluctuations, which are a common feature of the solar wind mostly due to various transients and disturbances, the solar wind in the heliosphere may also be reconstructed remotely by using interplanetary scintillation (IPS) data and/or white-light images. IPS measurements of a large number of radio sources are being used to reconstruct 3-dimensional (3D) velocity and density distributions of the solar wind by, e.g., the Computer Assisted Tomography (CAT) technique (Jackson et al., 1998, 2002, 2003, 2010; Manoharan, 2010), although the spatial and temporal resolutions are still not high enough to reflect fine structures of the solar wind.

The development of imaging techniques makes it possible for the solar wind to be imaged in a wide field of view (FOV) at a relatively high resolution in both space and time by space-borne white-light imaging devices. Such devices include, e.g., the Large Angle and Spectrometric Coronagraph (LASCO, Brueckner et al., 1995) on board the Solar and Heliospheric Observatory (SOHO), coronagraphs (COR-1 and COR-2) and heliospheric imagers (HI-1 and HI-2) in the Sun Earth Connection Coronal and Heliospheric Investigation (SECCHI) suite (Howard et al., 2008) on board the Solar Terrestrial Relations Observatory (STEREO) and the Solar Mass Ejection imager (SMEI, Eyles et al., 2003) on board the Coriolis satellite. According to the Thomson scattering theory (e.g., Vourlidas and Howard, 2006), the brightness of the K-corona is proportional to the number density of solar wind electrons, based on which the solar wind transients, or density inhomogeneous structures, can be revealed (e.g., Harrison et al., 2008).

To track the propagation of transients in white-light images, time-elongation maps, referred to as J-maps (e.g. Sheeley et al., 1999; Davies et al., 2009), are constructed. On J-maps of running-difference images, transients will leave bright and dark traces as they are moving outward. Based on different geometric models of transients and the assumption of constant outward velocity, several simple methods can be used to derive propagation (longitudinal) directions by fitting to the elongation-time profiles, using data from only one perspective, e.g., 'Point-P', 'Fixed-ϕ', 'Harmonic Mean' and 'Self Similar Expansion' methods (Howard et al., 2006; Kahler and Webb, 2007; Lugaz et al., 2009; Wood et al., 2009; Lugaz, 2010; Davies et al., 2012; Mostl and Davies, 2013). However, thanks to the dual perspectives achieved by the STEREO twin spacecraft, several triangulation methods have been developed to reconstruct the trajectories of various transients mostly without the need for the assumptions used by the geometric model techniques (Howard and Tappin, 2008; Koning et al., 2009; Lugaz et al., 2010). These methods were mostly applied to large-scale transients like CME/shock fronts, and need manual identification of the same features in both STEREO dual images. The small-scale transients, seen as 'blobs', appear in the white-light images and are detected as thin traces in the 'J-maps', especially clear at solar minimum (e.g., Sheeley et al., 2008, 2009). Rouillard et al. (2010) fitted the 'J-maps' of small transients during August and September 2007 by the 'Fixed- ϕ' method and find that they were frequently entrained within corotating interaction regions (CIRs). In the same way, they traced the small-scale transients not only appearing in either STEREO images but also recorded in in-situ observation at 1AU to find their solar origin (Rouillard et al., 2011).

Now we develop a new method to automatically recognize and locate solar wind inhomogeneous structures, especially transients, in the heliosphere based on the STEREO/HI dual imaging data. As the first paper in this series, here we focus on the transients along the Sun-Earth line and analyze the periodic behavior of the Earth-directed transients revealed by our method. The detailed description of the method is presented in the next section. The analysis of the periodic Earth-directed transients is given in Sec.3, 4 and 5. Finally, we provide our conclusions on the result and discussion about the method in section 6.

2. Data and Method

The two STEREO spacecraft (STEREO-A and STEREO-B) have been in a near 1 AU orbit of the Sun in the ecliptic plane since 2006. The SECCHI package on STEREO provides imaging observations, and consists of an extreme ultraviolet imager (EUVI, Wülser et al., 2004), two coronagraphs (COR-1 and COR-2, Howard et al., 2008), and the Heliospheric Imagers (HI-1 and HI-2, Eyles et al., 2009). Currently our method is developed based on HI-1 images because HI-1 has the desired FOV, cadence and signal-to-noise ratio (SNR) compared with other imagers as explained below. HI-1 and HI-2 image the outer corona and inner heliosphere in visible light near the ecliptic plane with an elongation from 4° to 24° and from about 18.7° to 88.7°, respectively. The size of HI-1 and HI-2's square FOV is 20° by 20° and 70° by 70°, respectively, while HI-1's FOV centered at 14° in elongation is closer to the Sun than that of HI-2's which is centered at 53.7° (Harrison et al., 2008). Such a wide FOV can help trace transients propagating in the inner heliosphere. The HI-1 imager cadence is 40 minutes while it is as long as 120 minutes for HI-2 (Eyles et al., 2009). With a stronger signal from scattered light from electrons nearer to the sun, compared to the background star-field and the Galaxy, the SNR of HI-1 image is higher than HI-2 and hence it is easier to observe transients in HI-1 images than in HI-2. Though COR-1 and COR-2 view the corona closer to the Sun, their FOVs are much smaller than HI-1's.

The separation angle between the Earth and each STEREO spacecraft increases about 22.5° per year (Kaiser et al., 2008). This separation angle was near 90° for both spacecraft from Jan 2010 to Feb 2011, as shown in Figure 1. This is the best time window for the twin STEREO/HI-1s to observe solar wind transients propagating along the Sun-Earth line. During that period, the Sun-Earth line was close to the two Thomson spheres (Vourlidas and Howard, 2006) in the FOVs of both HI-1s; the spheres denote the locations of maximum brightness per unit electron density along the line of sight (LOS).

The principle of our method is illustrated in Figure 2. Assuming that there is a featured object imaged by both HI-1s at the same time, we select a radial line, called the baseline, e.g., the Sun-Earth line here as indicated by the black dash line in Fig.2, and mark the segment of the baseline that crosses the featured object (as denoted by the thick lines) from the two HI-1s' images. If the object was on the baseline, the two segments are well overlapped (Fig.2a) and should be highly correlated in terms of the brightness recorded in the HI-1s' images. If the object was away from the baseline, the two segments should partially or even scarcely overlap (Fig.2b), and a low correlation could be expected. Based on this picture, we may calculate the linear correlation coefficient (cc) of two images' local intensity data to determine whether or not a featured object is located near the selected baseline. The obtained results will be compared with coronal observations from the STEREO/CORs and in-situ observations at 1 AU from the Wind spacecraft to justify the method in Sec.5. As mentioned before, here we only apply this method to the transients along the Sun-Earth line. However, we note that this method is not limited to a radial line; it can be easily extended to the whole heliosphere in 3D, which will be presented in the next paper.

We begin our method with STEREO/HI-1 Level 2 images (see Fig.3a and 3b), in which the 1 day background emissions, defined as the average of the lowest 25% of the data in a running window of 1 day on a pixel-by-pixel basis, have been removed (Eyles et al., 2009). Signals recorded in the HI-1 Level 2 images are a combination of the K-corona, F-corona and noise – mainly from the background star-field. The K-corona is photospheric light Thomson-scattered off solar wind electrons, including the contribution from solar transients. The F-corona is due to photospheric light scattered off dust. The F-corona evolves more slowly than the K-corona, and therefore can be removed by subtracting a previous image from the image of interest; this is a running difference technique that is often used for such data. The residual K-corona component on the running-difference images is just the contribution of transients. As for star-field, it slowly moves right in the FOV because of the STEREO's revolution around the Sun. To reduce the star-field as much as possible, we shift the previous image to match the FOV of the present one before performing the subtraction, a process we termed "shifted running-difference". HI-1 Level 2 data have a 40-minute cadence corresponding to a ~2-pixel shift, while the movement of solar wind transients in 40 minutes is typically over 20 pixels. Thus, the influence of such a shift on the signal is sufficiently weak. Then we apply a 3 x 3 median filter on the resultant difference images to remove residual spot-like noises. Compared with the original Level 2 HI-1 images (Fig.3a and 3b), the background star-field noise is reduced considerably (Fig.3c and 3d).

Knowing the position and FOV of the spacecraft and the baseline, we can project HI-1 images onto the meridian plane where the baseline lies based on temporal spatial positions of the two STEREO spacecraft in the Heliocentric Earth Ecliptic (HEE) coordination system. We use the latitude and distance from the solar center as polar coordinates to radially map the projected images and mark the spatial position of the transient indeed on this meridian plane (see Fig.3e and 3f). The resolution in latitude is set to 1°, while that in radial distance is 0.08-0.11 solar radii (Rs) varying with time to make sure the length in the radial distance is 1000 so the detail information can be kept in the projected images as much as possible from the HI-1 image with a size of 1024 × 1024. Then we compare the pixel values in the two projected HI-1 images near the selected baseline, i.e., the Sun-Earth line here. Here we set a 2D narrow box, which is 11-pixel (i.e., ±5°) wide in the latitude and 41-pixel (i.e., 3.2-4.4 Rs) long in the radial direction, running along the baseline in the dual images to calculate the correlation coefficient (cc) value between the two sets of the data points inside the same running box for two projected images. The cc we chosen is the linear Pearson correlation coefficient calculated by a formula of

$$cc = \frac{\sum_{i=1}^{n}(x_i-\bar{x})(y_i-\bar{y})}{\sqrt{\sum_{i=1}^{n}(x_i-\bar{x})^2}\sqrt{\sum_{i=1}^{n}(y_i-\bar{y})^2}} \qquad (1)$$

where x and y are two sets of data with a sample size of n. $\bar{x}$ and $\bar{y}$ are the average values of x and y. The value of cc is between -1 and 1, where 0 is no linear correlation, 1 and -1 are total positive and negative linear correlation. The larger positive cc, the better positive linear correlation. There is no uncertainty associated with a cc value, but it does have a confidence level, which depends on the size of data sets. The size of each data set in our method is 451 (11 in the latitude × 41 in the radial direction), that is large enough to guarantee a high confidence level. A distribution of the cc value along the baseline is therefore obtained as shown in Fig.3g. A region with cc ≥ 0.5 is treated as a correlated region, suggesting the presence of some inhomogeneous structures, including Earth-directed transients, whereas a low cc region, i.e., cc < 0.5, is interpreted as the absence of inhomogeneous structure along the baseline. Putting such cc distributions in a time sequence together, we get a cc map in the time-distance space as shown in Fig. 3h. In such a map, all Earth-directed transients can be recognized as the high-cc traces. The cc map looks similar to the traditional elongation-time 'J-map', except that we just track the possible transients existing on a preset radial line (i.e., the baseline) but not the whole transients in the FOV.

3. cc map of transients along the Sun-Earth line at large time scale

We apply the method to the HI-1 Level 2 data during the aforementioned time window from 1 Jan 2010 to 28 Feb 2011. The whole cc map setting the Sun-Earth line as the baseline during the 14 months is displayed in Fig.4a. The co-observed FOV of the two HI-1s during this period covers a radial distance from 19 Rs to 82 Rs along the Sun-Earth line. Because of the either or both missing HI-1 data at certain moments, there are data-gaps taking up 5.7% of the whole period of interest, displayed as black vertical stripes (see Fig.4a) in the cc map, which are filled with zeroes (see Fig.4b). On the cc map in Fig.4a, 54.3% of pixels out of data-gaps have a value between -0.1 and 0.1, hence zero is the most common choice for data-gaps. Since only the positive cc is meaningful in our study, we change the displayed range of cc to 0 ~ 1 to highlight the meaningful parts in Fig.4b. The longest data-gap is no more than one day and filling them with zeroes would not change the long-term features of cc.

The application of such cc maps on individual events will be discussed in the next paper. In the following sections, we focus on the long-term features of transients along the Sun-Earth line based on this 14 months' cc map to show the validity of the cc method. At first, we highlight long-term variations of correlated regions and reduce short-term noise. In the more than one-year period, most features appear too small to be visible. We further process the cc map to make them more visible at such a large time scale. As mentioned before, we use cc = 0.5 as the threshold to distinguish an Earth-directed transient. Thus, we may construct a binary cc map, in which the points with cc > 0.5 are set to unity and other points to zero, and smooth the binary cc map by replacing the value of each point with an average value over the 10-day time window centered on the point. Fig.4c shows the smoothed binary cc ($cc_p$ for short) map, in which large-scale features become prominent and noisy features are removed. Each pixel has a value ($cc_p$) between 0 and 1. It can be found that the pixel value generally decreases as distance increases, and there is a second peak pixel at most bins of time especially during the latter 7 months, of which the location moves from about 80 Rs in 2010 Feb to about 20 Rs in 2010 Dec (as denoted by the black solid line in Fig.4c). The most prominent region appears below 40 Rs during 2010 Oct - 2011 Jan. We check the solar cycle variation in terms of sunspot number, and find the different significance of the cc patterns is unrelated with it. Actually, these patterns not only result from continuous Earth-directed transients, but are also contributed by the following two effects.

3.1 Thomson scattering effect

According to Thomson scattering theory, (1) the intensity of the scattered light generally decreases rapidly with radial distance from the sun, and (2) the maximum scatter happens at the

Thomson sphere centered at the midpoint of the Sun-observer line with a diameter equal to the distance between the Sun and the observer (Vourlidas and Howard, 2006; Howard and Deforest, 2012). For the former phenomenon, the SNR decreases with the distance as well as the signal (i.e. scattered light) since the noise level remains almost the same. This effect is clearly seen in Figure 3. The intensity in the images (Fig.3e and 3f) and the cc value (Fig.3g) decrease with the distance. Due to the low SNR, we may miss weak transients at far distance. This is a reason why the pixel value generally decreases with the distance in the smoothed binary cc map.

For the latter phenomenon, a transient closer to the Thomson sphere is brighter, i.e., has a higher SNR, than the same transient at the same elongation angle but away from the Thomson sphere. If a transient is far away from the Thomson sphere of one spacecraft, its image would become blurry or noisy which may lead to a low cc value. In other words, if a transient locates on the baseline and near both Thomson spheres of the two STEREO spacecraft, its cc value will be high. The black dashed (dash-dotted) line in Fig.4c traces the cross-point of the Sun-Earth line and the Thomson sphere of STEREO-A (B). One can find that the aforementioned second peak pixels all locate between the two lines. As discussed by Howard and Deforest (2012), the Thomson sphere itself only marks the center of a wide region with high sensitivity for detection of heliospheric features by unpolarized heliospheric imagers like HI-1, hence the effect due to the distance from the Thomson sphere on the SNR may be not such prominent and there could be another effect that explains this second intensity peak, which we describe below.

3.2 Collinear effect

If a transient is located on the connecting line of the two spacecraft, i.e., they share the same LOS (as illustrated in Fig.5a), we have similar projected images of the transient even if it is located away from the baseline (Fig.5b), hence high cc values. In this case, a high cc value does not necessarily mean that a transient is propagating along the baseline. We termed this phenomenon as the collinear effect.

The black solid line in Fig.4c traces where the Sun-Earth line crosses the line connecting the two STEREO spacecraft, which is right between the two black dashed/dash-dotted lines. The second peak pixels locate near the line, consistent with the collinear effect. During the November and December 2010 the three black lines converge toward the inner boundary of the FOV, where the features are most prominent and the cc values in this region may be overestimated compared to other regions due to the above effects.

4. Periodicity of Earth-directed transients

In the smoothed binary cc map (Fig.4c), we can see cc values peak over a period of about 27 days. This pattern is not clearly visible in the original cc map (Fig.4b). To reveal the hidden periodicities, we applied Fourier transform to the time series of cc values in the cc map at each distance, and obtained a power spectrum (P) in the frequency domain. The statistical significance of Fourier spectral power peaks is evaluated by calculating the variance, $\sigma^2$, of the randomized time series of the cc values, which is assumed to be Gaussian-distributed white noise (see Sec.2.1 of Lou et al., 2003). Here we consider that a peak in the Fourier frequency domain is statistically significant if its power exceeds $5\sigma^2$. Figure 6a shows the square root of the power spectrum of the cc map, which has been normalized by $\sigma$. For clarity, the horizontal axis shows the period instead of frequency. In this figure, we can find a very significant peak region at the period of near 27 days over the distance of less than 50 Rs. Besides, this periodic structure disappears in the map of the earlier 7 months (Fig.6b) but is much clear in the later 7 months (Fig.6c). This is because in the later 7 months, not only where the Sun-Earth line crosses the line connecting the two STEREO spacecraft (see black solid line in Fig.4c) but the cross-points of the Thomson spheres and the Sun-Earth line (see black dashed line and black dash-dotted line in Fig.4c) are closer to the high SNR region (i.e. the region near the solar center). Other peaks greater than $5\sigma$ are much weaker, such as the regions with periods of integral multiples of about 7 days (e.g. 7-9 days, 14-15 days, 21 days) which have the similar features as the 27-day periodicity discussed above and even the period of 27 days itself is about 4 times of 7 days, so they may all be related to the solar self-rotation (with a period of just 27 days). It should be noted that in spite that some unreliable features may appear in the cc map because of the aforementioned effects, they cannot produce fake periodicities. In the following analysis, we focus on the 27-day periodicity and check its source.

To explore which kind of transients, i.e., large-scale ones, like CMEs and shocks, or small-scale ones, contribute to the periodic phenomenon, we identify CMEs and shocks on the cc map based on the manual STEREO/HI-1 CME catalogue (HICAT) (https://www.helcats-fp7.eu/catalogues/wp2_cat.html), a product of the HELiospheric Cataloguing, Analysis and Techniques Service (HELCATS) project (Harrison et al., 2018). The CME events in HICAT were all extended over 20° to be distinguishable from blob-like small features. The CMEs we chose from HICAT as large-scale transients have to fulfill two conditions: 1) observed by HI instruments on both STEREO-A and B, and 2) the CME spans over the position angle of the Sun-Earth line in the FOVs of STEREO-A and B. After CMEs have been identified, we suppose any other transients are small-scale ones. According to this classification scheme, we can judge which category a point on the cc map belongs to based on corresponding projected images like Fig.3e, 3f. Then we split cc map into $cc_L$ and $cc_S$ maps for large and small-scale transients, respectively, in which the values of all the pixels not belonging to $cc_L$ or $cc_S$ map are set to 0. To display the long-term features of the two kinds of transients, we finally use $cc_L/cc_S$ to construct smoothed binary cc for large/small-scale transients named as $cc_{p,L}/cc_{p,S}$ in a similar way as constructing $cc_p$ from cc. It is noted that $cc_{p,L} + cc_{p,S} = cc_p$. In Figure 7, we display $cc_p$, $cc_{p,L}$ and $cc_{p,S}$ at the distance of 20 Rs (red lines), 30 Rs (orange lines) and 40 Rs (green lines) from 15 Aug 2010 to 25 Dec 2010. The ratio of averaged $cc_{p,S}$ to averaged $cc_{p,L}$ is about 2.17, 1.05 and 0.62 at the distance of 20 Rs, 30 Rs and 40 Rs, respectively. Please note that the ratios here in fact reflect the ratio of occupation times of the two kinds of transients at certain distance along the Sun-Earth line. For small-scale transients, many of them are not on or near the Sun-Earth line but either out of the ecliptic plane or on the ecliptic plane but in different directions. For large-scale transients they have higher probability to span over the Sun-Earth line and last for a longer time than small ones. Thus, the real ratio of the numbers of the two kinds of transients at a certain distance should be much higher. From these ratios, we can see that the small-scale transients in the solar wind are more frequent than large-scale transients; but with the increasing distance, the mass increase in large-scale transients like CMEs (Lugaz et al., 2005; DeForest et al., 2013; Feng et al., 2015) can keep their SNR high enough to be clearly observed, while the rapidly reduced SNR makes the small-scale transients become much weaker than large-scale transients, and therefore the ratio decreases significantly. This also suggests that most small-scale transients look like merged and diffused into background solar wind in white-light images beyond 40 Rs.

Regarding to the 27-day periodicity, we find that the periodicity of small-scale transients is much clearer than that of large-scale transients, especially during Nov and Dec 2010 in Figure 7c. This is displayed in Figure 6a and 6c as well. We find the peak region at the period of 27 days at the distance bellow 50 Rs where large-scale transients dominate but less than 50 Rs where small-scale transients can be seen. This is reasonable because most notable small-scale transients near the ecliptic plane come from coronal streamers above solar active regions (Plotnikov et al., 2016), which may live for as long as several Carrington rotations. A more detailed analysis of the periodic transients is given in the next section.

5. Observations along the Sun-Earth line

To learn more about the small-scale transients and in turn confirm the results of the cc map constructed by our method, we use STEREO-COR2 imaging observations (Howard et al., 2008) to check the density variation at the distance of 5 Rs to trace the source of the small-scale transients, and use Wind spacecraft (Ogilvie et al., 1995) near the Earth to check the in-situ solar wind conditions.

5.1 Observations at 5Rs

STEREO-COR2 images reveal bright coronal streamers - large-scale magnetic structures that are closed at the base and open to the heliosphere above, that are readily imaged due to the enhanced plasma density. Occasionally, we can observe CMEs as bright expanding structures. We apply the same projection process on STEREO COR-2 images as we did to the STEREO HI-1 images, choosing the meridian plane parallel to the Sun-Earth line as the projection plane. On this plane, we sample the brightness data along the latitudinal direction from -80° to 80° at the distance of 5 Rs from 15 Aug 2010 to 25 Dec 2010. Then we normalize the intensity with equation (2).

$$I_n = \frac{I - \langle I \rangle}{\sigma_I} \qquad (2)$$

Here $I$ is the intensity, $\langle I \rangle$ and $\sigma_I$ are the average value and standard error of $I$. Finally, we display time-latitude maps of $I_n$ for STEREO-A/B in Figure 8a and 8b. From 15 Aug 2010 to 25 Dec 2010, the separation angle between the Earth and the two STEREO spacecraft is near 90°, hence the Sun-Earth line is near the Thomson spheres of both STEREO spacecraft in the FOV of COR2. Thus, the main structures on the two maps are on or near the selected meridian plane.

On the maps, the bright vertical stripes represent CMEs and/or shocks while the bright prolonged curves or plumes are coronal streamers. The large-scale dim regions are coronal holes. During this period of time, there are five Carrington Rotations, CR2100 to CR2104. Near the ecliptic plane we find five main coronal holes and five streamer groups, which all appeared once in one Carrington Rotation. We plot $\overline{I_n}$, the average $I_n$ from -5° to 5° in Figure 8c. The valleys (coronal holes) and peaks (streamer groups) appear alternately with a period of about a month which is associated with the 27-day periodicity of small-scale transients found in the cc map. Besides, we find that the CMEs (simply taking the midpoints of the vertical stripes) are distributed at a wide range in latitude near the streamers but far away from coronal holes in Figure 8a and 8b. Many CMEs on the ecliptic plane displayed as spikes in Figure 8c appear not to originate near the local streamer belts but originate at different latitudes and span across or expand into the ecliptic plane, which explains why we can find these spikes not only during the peaks (streamer groups) but also the valleys (coronal holes) of $\overline{I_n}$ in Figure 8c.

5.2 Observations near the Earth

Figure 9e and 9g show the proton number density and solar wind bulk speed observed by the Wind spacecraft from 15 Aug 2010 to 25 Dec 2010. Since the cadence of HI-1 images is 40 min, the in-situ observational data at 1AU are averaged every 40 min. During this period, there was no interplanetary coronal mass ejection (ICME) passing over the Earth according to the catalog of ICMEs established based on the in-situ data from the Wind and ACE spacecraft by Chi et al. (2016). That means most structures in the in-situ observations near the Earth during that period can be related to small-scale transients.

We find that fast solar wind (defined with the speed > 500 km s$^{-1}$) arrives at 1AU about every month. So we mark the fast solar wind with shadow regions in Figure 9e-g and label them f1 through f5. Between any two consecutive f regions, there is a slow wind region (marked as s1 through s4). In imaging data, a transient is visible because of the inhomogeneity in density. Thus, we calculate the mean proton number density, $\bar{n}_p$, and the mean value of the absolute time gradient of proton number density, $\Delta n_p$, in each f or s region as shown in Figure 9f. The latter parameter represents the inhomogeneous level of the proton density in each region. We find that $\bar{n}_p$ is less than 4 cm$^{-3}$ and $\Delta n_p$ less than 0.4 cm$^{-3}$ in the fast solar wind, while the situation is very different for slow solar wind, suggesting that the detected small-scale transients in the cc map should correspond to the slow solar wind. The high relevance between small-scale transients at 1 AU and slow solar wind was confirmed by previous statistical studies (Feng et al., 2008; Yu et al., 2016). It is consistent with the present result that the small-scale transients mostly come from coronal streamers where the slow solar wind originates.

5.3 Multi-distance analysis of small-scale transients along the Sun-Earth line

For convenience, Figure 9a is the same as Figure 8c representing observations at 5 Rs, and Figure 9b-d show the profiles of cc$_{p,s}$ at the distance of 20Rs, 30Rs and 40Rs, respectively, which have been displayed in Fig.7d. Assuming the speed of the interfaces between the fast and slow solar winds as marked by the vertical dash lines in Figure 9g remains at 500 km s$^{-1}$ from 5 Rs to 1 AU, we may estimate their arrival times at different distances, which have been indicated by dash lines in Figure 9a-d. The shadow/white regions mean expected fast/slow solar wind regions. We find that these regions are consistent fairly well at different distances. This result further confirms that the small-scale transients come from coronal streamers and mainly contribute to the 27-day periodicity in the cc map.

6. Summary and Discussion

In this work, we proposed a new method to recognize and locate solar wind transients by combining two simultaneous STEREO-A and B's HI-1 images, in which a cc map is constructed to identify density disturbances in the solar wind. Different from existing methods for the reconstruction of solar transients, this method does neither presume any geometry of transients, nor the propagation direction. In this way, we can deal with a huge number of dual images in time sequence automatically as we have done to the 14-month HI-1 data in this paper. By applying the method to the solar wind propagating along the Sun-Earth line, the constructed cc map manifests Earth-directed transients directly without involving tracing them in the HI-1 images or J-maps manually. Since the value of cc between two series of data does not change even if either of the two series is scaled upward or downward as a whole, our method shows the merit that a high cc value can be obtained as long as the SNRs of transients are sufficiently high.

However, the new method has some weaknesses as well, e.g., the collinear effect. In principle, we can determine a transient's longitude easily by changing the longitude of the baseline and comparing their cc values. But if a transient is aligned along the line connecting both spacecraft, the cc will be always high no matter which longitude is chosen. In this case the information of the transient's propagating direction is lost. Secondly, it is a highly ideal assumption that the source electrons concentrate in a relatively small volume along the LOS. This assumption may be good enough for blob-like small-scale transients, but not for large-scale and complicated transients, in which there might be different electron distributions along different lines of sight for different observers leading to very different white-light images of the same structure (i.e., low cc). Third, if two similar transients propagate one by one, we might still find a high cc value even if these transients are far away from the baseline, because the calculated cc may be between the first transient on one image and the second transient on the other image if they are projected to the same position on the baseline. These weaknesses may be improved by involving imaging data from more perspectives, e.g., the images from SOHO/LASCO or other heliospheric white-light images from future spacecraft.

Using this method, we show the features of solar transients along the Sun-Earth line in the ascending phase of solar cycle 24 from Jan 2010 to Feb 2011. A 27-day period is found in the constructed time-distance cc map. By analyzing the coronagraph observations at 5 Rs and in-situ measurements of the solar wind speed and density near the Earth, we conclude that the cc map reveals a real 27-day periodic pattern of solar transients, which is attributed to the small-scale transients originating from coronal streamers and propagating as slow solar wind. Furthermore, it is suggested by the cc map that small-scale transients are more frequently present than large-scale transients (Fig.7c-d) by a factor of at least 2, and quickly diffused into background solar wind within about 40 Rs. Again, it should be noted that the ratio is underestimated, because our method only traces the transients near the Sun-Earth line. Large-scale transients, like CMEs, have a much wider span angle than small-scale transients, and may cross the Sun-Earth line even though the apex of them are off the Sun-Earth line. Small-scale transients must be about on the Sun-Earth line to be detected by our method.

Rouillard et al. (2010) fitted the tracks in the HI-1 J-map during August and September 2007 and found small-scale transients in solar minimum were frequently entrained by CIRs. This is because the fast wind from coronal holes at high longitude can spread to low latitude in any longitude and may overtake small-scale transients in the slow solar wind near the ecliptic plane and form a CIR. At solar maximum of solar cycle 24, coronal holes extend to the equator and the number of CMEs is large. Meanwhile, small-scale transients at the ecliptic plane are rare in contrast to the solar minimum (Yu et al., 2016). Sanchez-Diaz et al. (2017) just found groups of small-scale transients in the slow wind originating along a north-south oriented neutral sheet near a coronal hole at the equator in 2013 during solar maximum. This north-south oriented neutral sheet is thin, hence the corresponding small-scale transients were still overtaken by fast wind from the near coronal hole and ended with a CIR. In the ascending phase, we find the behavior of small-scale transients is different from the minimum and the maximum according to our example. Most of the small transients at 1 AU in Figure 9 are widespread in the slow wind without ending with a CIR as the cases at solar minimum or maximum. In Figure 8, we find that coronal holes spread to the equator as well but only from one side, and the other side is a group of streamers. The orientations of these streamer stripes are mostly in northwest-southeast direction. Thus, these streamers in the ecliptic plane are still thick, and only a few small-scale transients originating near a coronal hole can be captured by fast wind and end up as CIRs. The others can propagate to 1AU in slow wind without entrained by fast wind.

Adopting multiple baselines on the same meridian plane but at different latitudes or on meridian planes at different longitudes, our method has the potential of revealing the 3D solar wind structures in the heliosphere. This will be presented in the next paper. Besides, we choose the threshold of cc = 0.5 just by experience, but we have tried to change this from 0.25 to 0.75 and find the same 27-day periodic pattern in the corresponding $cc_p$ map. We will refine the value of the threshold in our future work.

**Acknowledgments**   The STEREO/SECCHI data are produced by a consortium of NRL (USA), RAL (UK), LMSAL (USA), GSFC (USA), MPS (Germany), CSL (Belgium), IOTA (France), and IAS (France). The SECCHI data presented in this paper were obtained from STEREO Science Center (https://stereo-ssc.nascom.nasa.gov/data/ins_data/secchi_hi/L2/). The Wind data were obtained from the Space Physics Data Facility (https://cdaweb.sci.gsfc.nasa.gov/). This work is supported by the grants from NSFC (41574165, 41774178 and 41761134088) and the fundamental research funds for the central universities (WK2080000077).

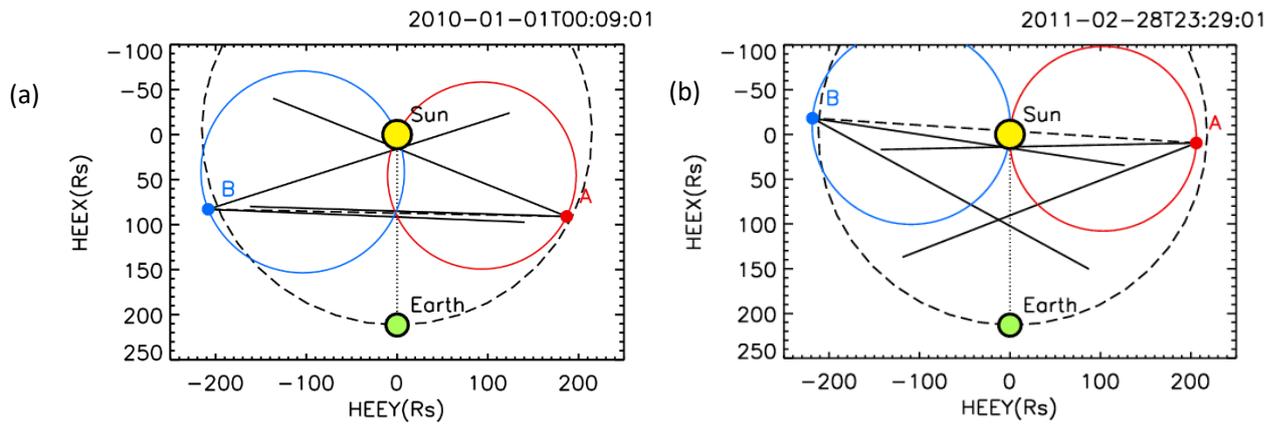

Figure 1. (a) Positions of STEREO-A/B, Earth and the Sun on the ecliptic plane at 00:09:01 UT on 1 Jan 2010. (b) The same as (a) but at 23:29:01 UT on 28 Feb on 2011. The black dashed circle line represents the Earth's orbit. Black solid lines delimit the angular extents of FOVs of both HI-1 cameras in the ecliptic plane. The Sun-Earth line is shown as a dotted line. The intersection of STEREO-A/B's Thomson sphere and the ecliptic plane is shown as the red/blue solid circle.

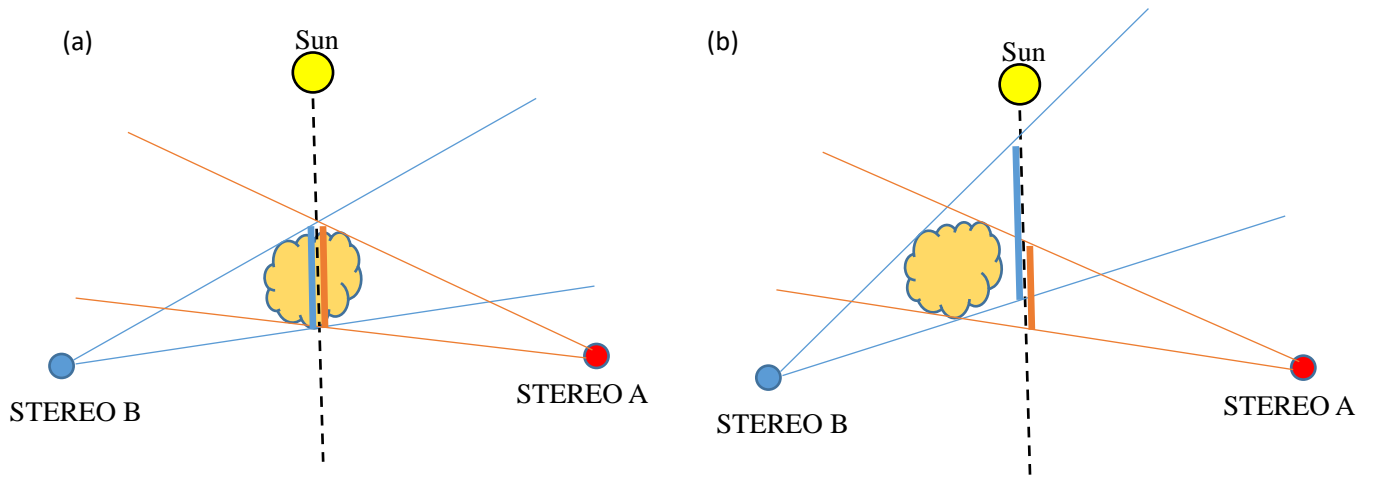

Figure 2. The schematic diagram of the CI method. The yellow, blue/red points are marked as the Sun, STEREO-B/A spacecraft. The black dashed line with one end on the Sun is the baseline. The orange cloud is the featured object in the solar wind. The two thin blue/red radial solid lines with an end on STEREO-B/A are the lines of sight that meet this object exactly at a tangent and the thick blue/red solid line indicates the segment of the baseline covering the featured object from the HI-1s' images. (a) The case when the featured object is on the baseline. (b) The case when this featured object is away from the baseline.

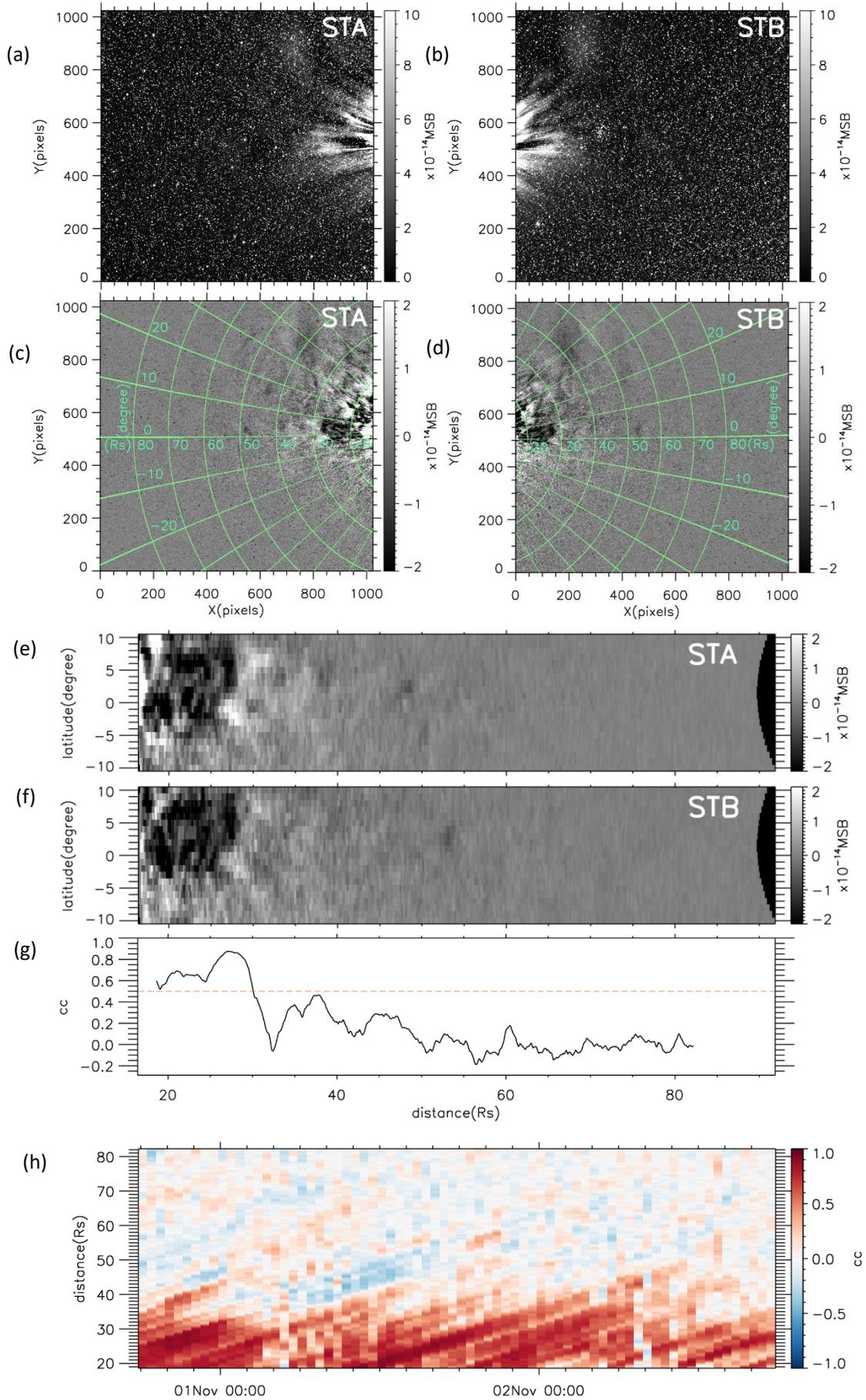

Figure 3. For panels (a)-(d), the left/right hand frames are for HI-1A/HI-1B. (a) and (b) The HI-1 Level 2 data of STEREO-A/B at 18:09 UT on 1 Nov 2010. (c) and (d) The shifted running-difference images of (a) and (b). The green lines are contours of latitude and distance from the solar center on the meridian plane parallel to the baseline (i.e., the Sun-Earth line). (e) and (f) The projected (c) and (d) sectors within the latitude of +/- 10° in the distance-latitude space along the baseline, i.e., the Sun-Earth line. (g) The cc value along the baseline. See section 2 for the definition and details of the cc calculation. The red dashed line marks cc = 0.5. (h) The cc map in the time-distance space from 18:09 UT on 31 Oct 2010 to 18:09 UT on 2 Nov 2010.

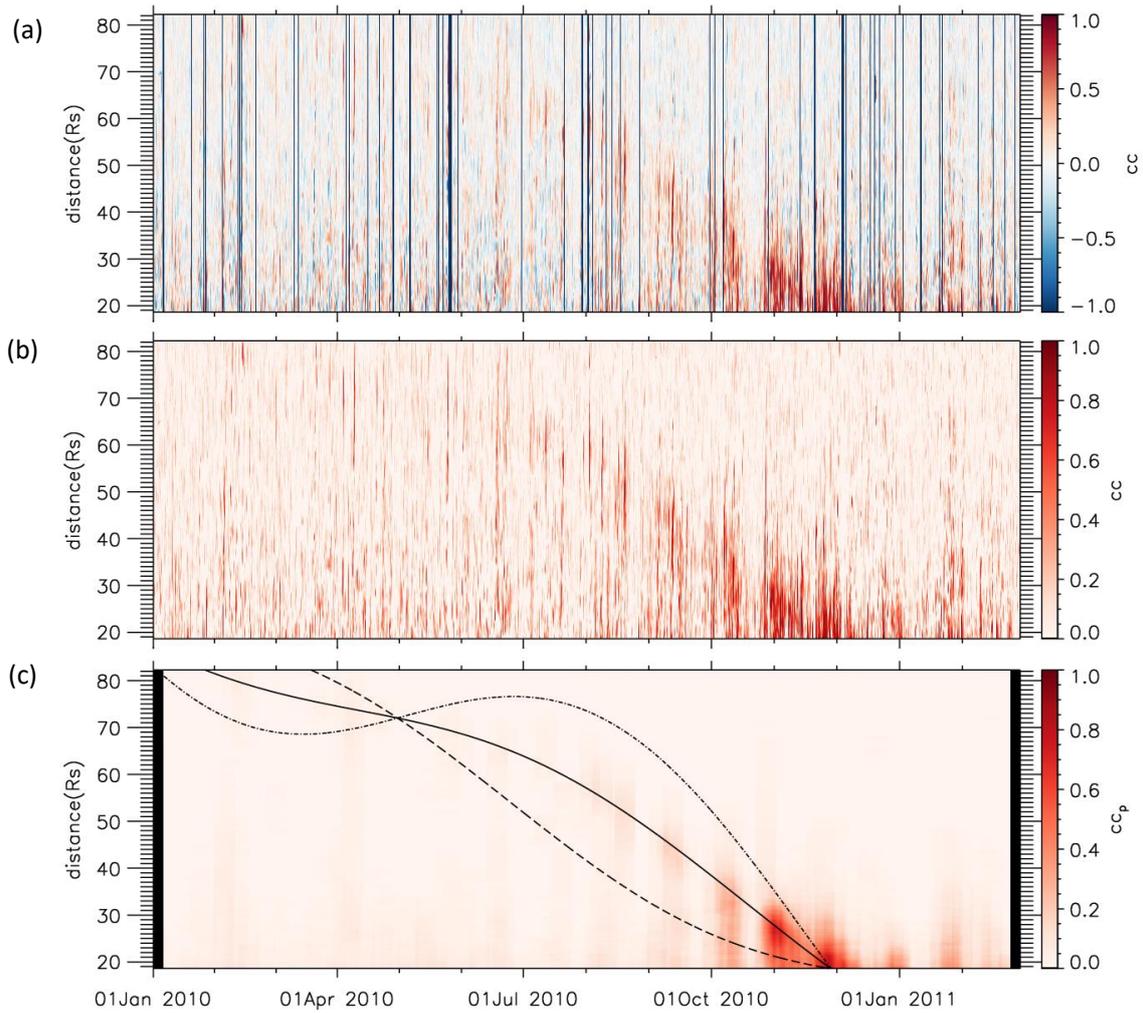

Figure 4. (a) The cc map in time-distance space from 1 Jan 2010 to 28 Feb 2011 along the Sun-Earth line. The black stripes are data-gaps because either or both HI-1 data are missing at these times. (b) The positive cc map changing the displayed cc range after filling data-gaps with zeros. (c) The smoothed binary cc map from 1 Jan 2010 to 28 Feb 2011 along the Sun-Earth line. The black solid line traces the cross-point of the Sun-Earth line and the connecting line between the two STEREO spacecraft. The black dashed line traces the cross-point of the Sun-Earth line and the Thomson sphere of STEREO-A, and the black dash-dotted line traces the cross-point of the Sun-Earth line and the Thomson sphere of STEREO-B.

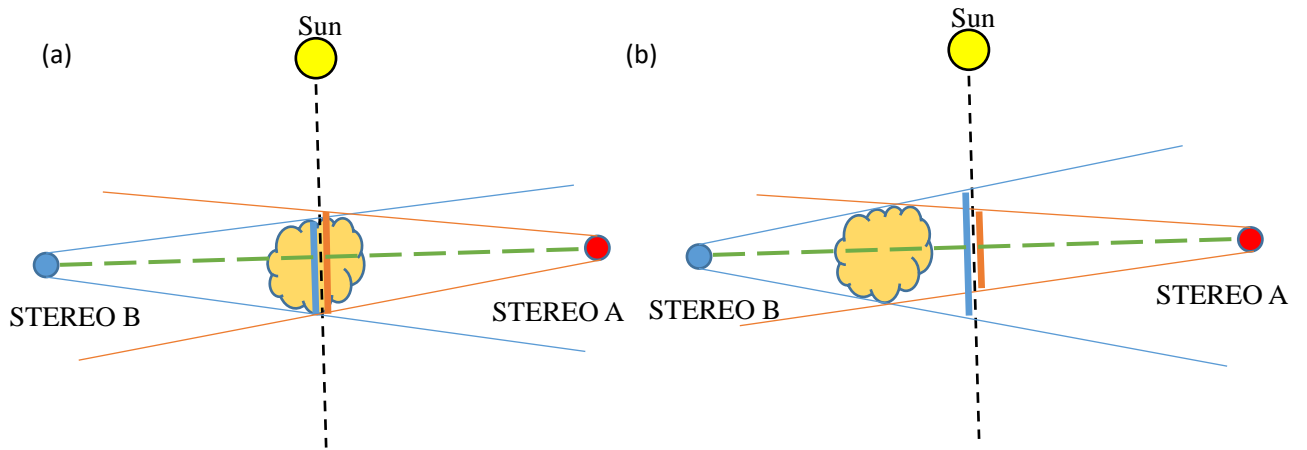

Figure 5. The schematic diagram of the collinear effect. The green dashed line is the connecting line between the two STEREO spacecraft. (a) The case when the featured object is on the baseline. (b) The case when this featured object is away from the baseline.

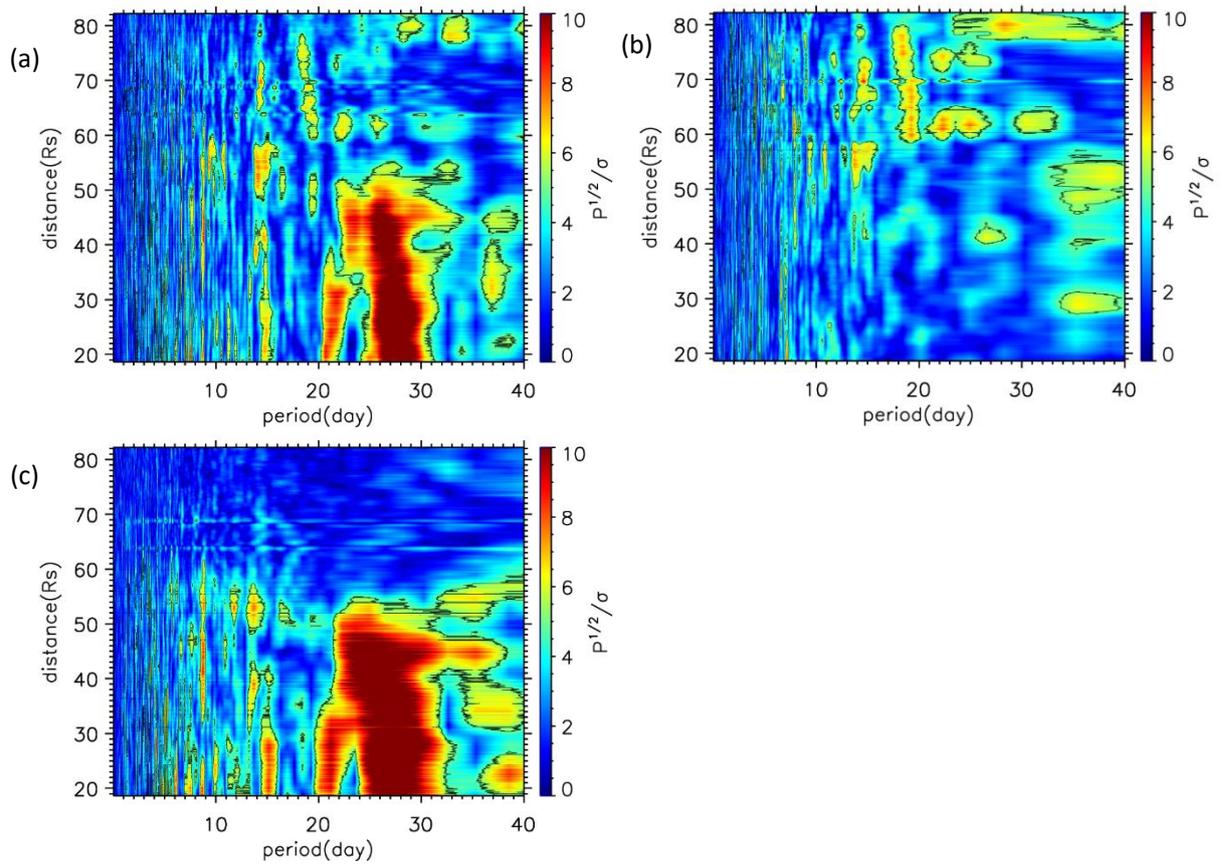

Figure 6. (a) The square root of the power spectrum of the cc map for the period from 1 Jan 2010 to 28 Feb 2011, which has been normalized by σ. (b) The same but for the first 7 months from 1 Jan 2010 to 31 Jul 2010. (c) The same but for the next 7 months from 1 Aug 2010 to 28 Feb 2011.

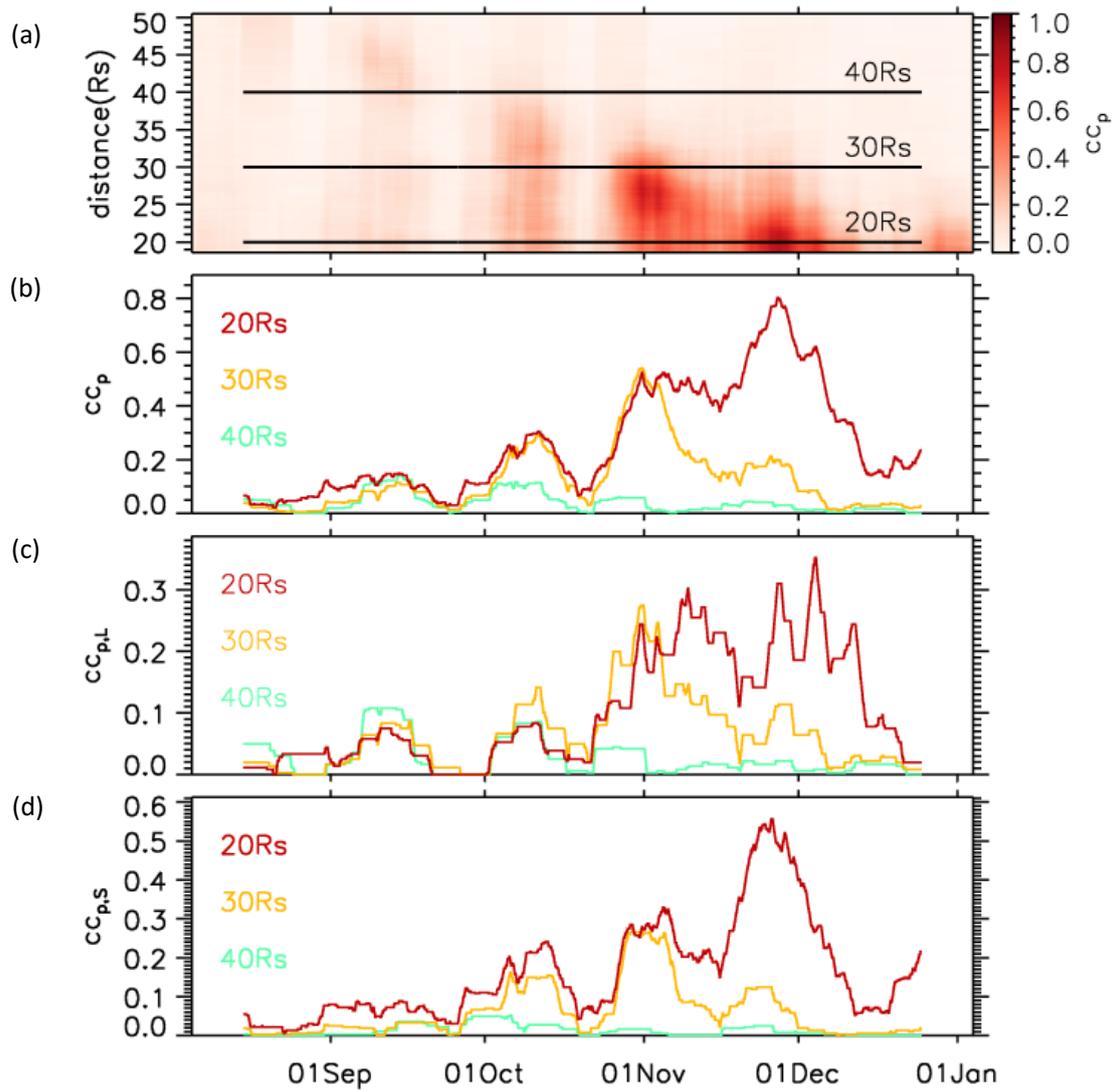

Figure 7. (a) The smoothed binary cc map in time-distance space from 15 Aug 2010 to 25 Dec 2010 along the Sun-Earth line. The black solid lines mark the distance of 20, 30 and 40 Rs. (b), (c) and (d) The cc values of all transients ($cc_p$), large-scale transients ($cc_{p,L}$) and small-scale transients ($cc_{p,S}$) at the distance of 20 Rs, 30 Rs and 40 Rs from 15 Aug 2010 to 25 Dec 2010. The red, orange and green lines represent the cc values at the distance of 20 Rs, 30 Rs and 40 Rs, respectively.

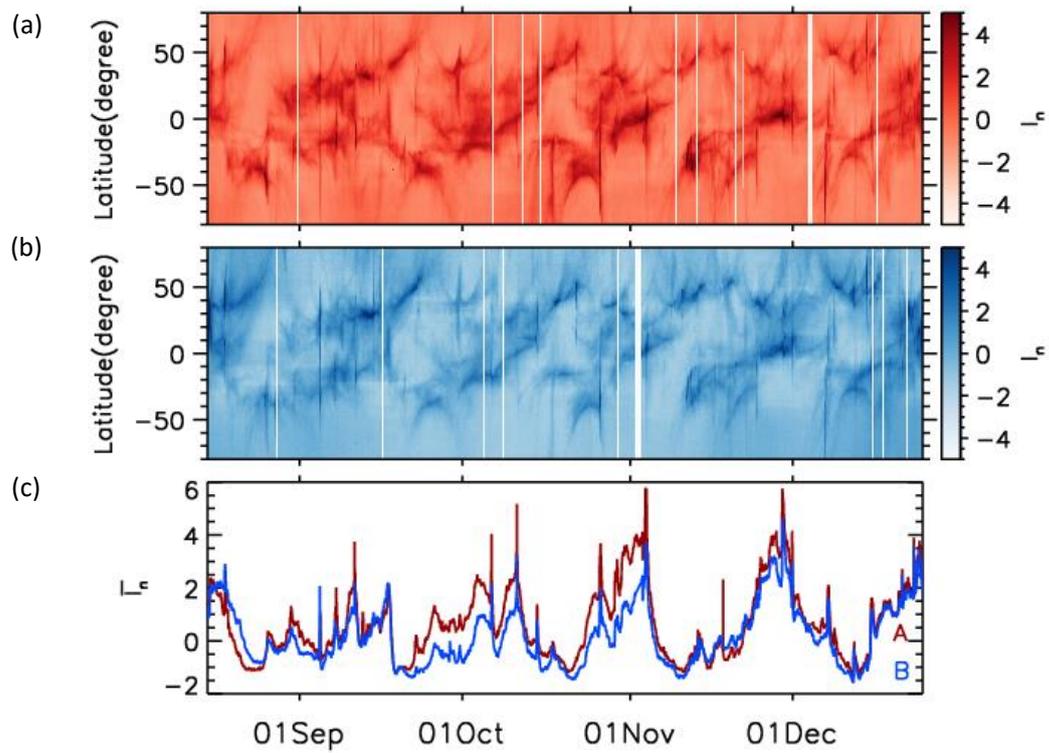

Figure 8. (a) Time-latitude maps of the normalized intensity, $I_n$, at the distance of 5 Rs from solar center on the meridian plane parallel to the Sun-Earth line for STEREO-A. About the definition of $I_n$, please see Section 5.1 for explanation. (b) The same as (a) except it is for STEREO-B. (c) $\bar{I}_n$, the averaged $I_n$ from -5° to 5°. The red/blue corresponds to observations from STEREO-A/B.

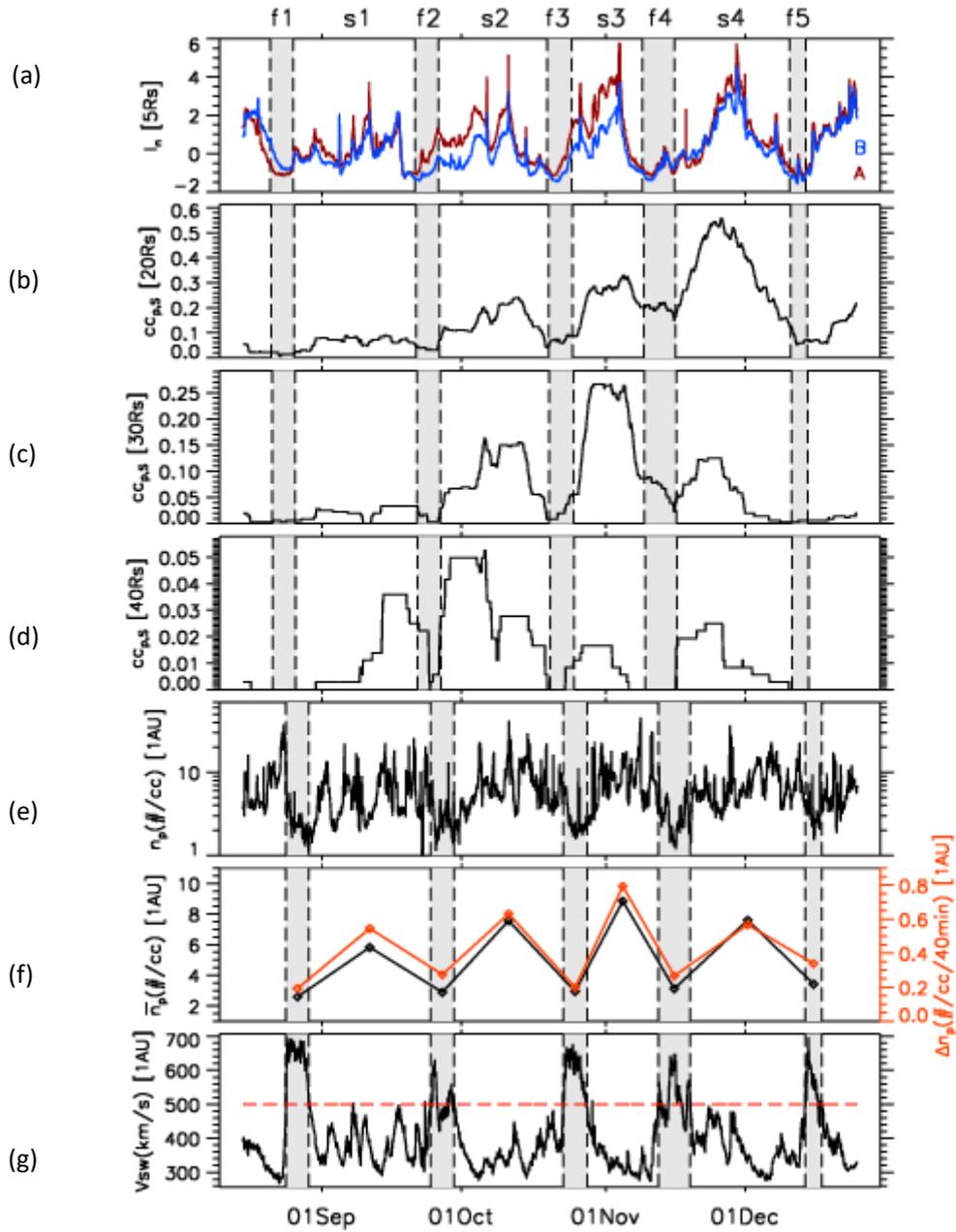

Figure 9. (a) The same as figure 8c. (b), (c) and (d) The value of cc$_{p,S}$ at the distance of 20 Rs, 30 Rs and 40 Rs from 15 Aug 2010 to 25 Dec 2010. (e) Solar wind proton number density at 1 AU detected by Wind. (f) The mean proton number density ($\bar{n}_p$) and mean absolute central difference of proton number density ($\Delta n_p$) of each region f and s. (g) Solar wind bulk speed at 1 AU detected by Wind. The red dash line in (g) represents the speed of 500 km/s. The black vertical dash lines are marked as the time when solar wind speed is 500 km/s. The shadow/white regions mean expected fast/slow solar wind regions, called f1-f5/s1-s4.